# Conservation of Helium while Maintaining High System Purity

**M. White, J. Theilacker, M. Barba**

Fermi National Accelerator Laboratory, PO Box 500, Batavia, IL 60510

Email: mjwhite@fnal.gov

**Abstract**. Recent helium shortages and helium price increases have lead to an increased emphasis being placed on conserving helium. The need to conserve helium must be balanced with need to maintain the high levels of purity necessary to prevent operational problems caused by contamination. Helium losses and contamination control are especially important for test stands that have cryogenic distribution systems operating continuously with frequent changeover of cryogenic temperature components that are being tested. This paper describes a mathematical model to estimate the quantity of helium lost and the purity of the helium after the pump and backfill procedure is complete. The process to determine the optimal time during pump down to cut off pumping and start backfilling is described. There is a tradeoff between trying to achieve the lowest possible pressure during pumping and the quantity of air leaking into the volume while pumping is occurring. An additional benefit of careful selection of pump and backfill parameters in conjunction with real-time pressure monitoring can reduce the labor and time required to complete a successful pump and backfill procedure. This paper is intended to be a tool for engineers to review their pump and backfill procedures and measured data to optimize helium losses, system purity, and labor required.

## 1. Introduction

Helium is a non-renewable resource and eventually known helium reserves will be depleted. This suggests that helium shortages and price increases will be more severe in the future than they have been in the past. Helium is a byproduct of uranium and thorium decay. Commercially available helium is typically derived from natural gas, since helium released during radioactive decay can be trapped in the same natural formations as natural gas [1]. Reduced fossil fuel extraction to mitigate climate change could also result in a reduced supply of helium. Only a few locations in the United States have natural gas deposits with >= 0.3% helium. Dilute concentrations make separating helium expensive. Limited number of locations where extracting helium is economically feasible. Helium is an international commodity and geopolitical conflict can lead to helium shortages and/or price spikes. There are a limited number of helium supply facilities and distributors. Force majeure facility closure of a single facility can cause restrictions on quantities of helium being delivered over a wide area.

Finding ways to conserve helium now starts the accrual of financial savings and conserves a finite resource. Reducing helium losses now also mitigates the impact of future helium price increases and/or shortages.

One method of reducing helium losses is to carefully evaluate pump & backfill procedures to achieve desired purity levels with the minimum amount of helium loss. Clean-up of a helium cryogenic system is accomplished by vacuum pumping the gas, initially air or nitrogen, out of the system and backfilling with helium. Repeated pump and backfill cycles leave the helium pure enough to connect the volume to a helium system. Unnecessary backfill cycles are a helium loss that can be avoided.

A dry nitrogen purge is typically applied prior to pump and backfilling to remove residual water. The first reason is that pumping is not very effective at removing water, since evaporation cools the water and limits the vapor pressure. The second reason is that the presence of water can limit the ultimate pressure reached during the pumping cycle. At a temperature of 300 K the vapor pressure of water is 35 mbar. If the ultimate pressure is limited to 35 mbar, then additional backfill cycles will be needed and unnecessary helium losses will be incurred

## 2. Helium Purity

Multiple cycles are required to get down to a contamination level acceptable as an input to a $LN_2$ temperature charcoal adsorber (< 50 $PPM_v$). More stringent purity requirements may apply if the helium used to sweep the volume after pump and backfill cannot be fully routed to $LN_2$ temperature charcoal adsorber. The final purity of an ideal system with no leakage after pump and backfilling can be estimated using equation (1)

$$\theta_{Ideal} = \left(\frac{P_{end}}{P_{start}}\right)^N x\ 10^6 \qquad (1)$$

where $\theta_{Ideal}$ is the contamination concentration in $PPM_v$ on a system with no leaks, $P_{start}$ is the starting pressure of the pump cycle, generally atmospheric pressure, $P_{end}$ is the ending pressure of the pump cycle, and $N$ is the number of pump and backfill cycles. As shown in Table 1, with an end pressure of 150 mbar it takes 6 cycles and 51.1 $m^3$ of helium to clean a 10 $m^3$ volume to less than 50 $PPM_v$. Reducing the end pressure to 50 mbar decreases the number of cycles to 4 and the 38 $m^3$ of helium used. Reducing the end pressure to 5 mbar further reduces the number of cycles to 2 and the helium used to 19.9 $m^3$. Tightening the end pressure requirement from 150 mbar to 5 mbar reduces pump and backfill helium usage by 60% while achieving the same purity. Schedule and labor costs also improve due to the reduced overall pump and backfill procedure time required.

**Table 1.** Expected contamination level after each pump and back fill cycle using equation (1). The pumped volume was arbitrarily selected as 10 $m^3$.

| Cycle | $P_{end}$ = 150 mbar | | $P_{end}$ = 50 mbar | | $P_{end}$ = 5 mbar | |
|---|---|---|---|---|---|---|
| | Purity $PPM_v$ | Helium $m^3$ | Purity PPMv | Helium $m^3$ | Purity $PPM_v$ | Helium $m^3$ |
| 1 | 148,075 | 8.5 | 49,358 | 9.5 | 4,936 | 10.0 |
| 2 | 21,926 | 17.0 | 2,436 | 19.0 | 24.4 | 19.9 |
| 3 | 3,247 | 25.6 | 120 | 28.5 | 0.1 | 29.9 |
| 4 | 481 | 34.1 | 5.9 | 38.0 | 0.0 | 39.8 |
| 5 | 71.2 | 42.6 | 0.3 | 47.5 | 0.0 | 49.8 |
| 6 | 10.5 | 51.1 | 0.0 | 57.0 | 0.0 | 59.7 |

## 3. Determining Pump and Backfill Parameters

The purpose of this paper is to provide a methodology to quantitively optimize pump and backfill parameters to achieve the desired purity level with a minimum loss of helium and a minimum time required to complete the procedure. Using the simple formula in equation 1 is insufficient by itself, since real systems have leaks and other limitations on ultimate pumping pressure. Reviewing cryogenic literature and cryogenic engineering handbooks yielded little useful information on optimizing pump and backfill parameters using calculations and experimental data. Furthermore, different people at different times and locations have selected various pressure and time criteria to stop pumping and various numbers of pump and backfill cycles even within a single organization such as Fermilab. The methodology in this paper will help organizations standardize their pump and backfill procedures and reduce the associated helium losses.

A critical component necessary for optimizing pump and backfill procedures is the pressure transmitter. The pressure transmitter must be capable of reading accurately with sufficient precision at the desired vacuum level and should be on a periodic calibration. Additionally, transmitter readings should be read through a control system where the reading can be plotted in real-time on a semi-log plot. Ideally the transmitter is placed at the opposite end of the pumped volume from the vacuum pump for continuous accurate readings.

If it is unavoidable to place the transmitter on the pumping line, then there needs to be an isolation valve between the pump and the transmitter. The drawback is that the isolation valve will need to be periodically closed and enough time must be given for the vacuum pressure to equalize across the pumped volume. Therefore, the procedure will take longer, more time at vacuum pressures provides more time for air to leak in, and it becomes more difficult to determine when pumpdown becomes non-linear on a semi-log plot.

The pressure versus time plot on a semi-log scale is ideally a straight line, so having the capability to monitor the pressure transmitter signal in real time is helpful in optimizing when to stop the pumpdown. The pumpdown should be stopped once the pumpdown curve becomes non-linear, since at this point the quantity of air leaking into the system becomes significant relative to the quantity of air or air/helium mixture being pumped out.

In addition to air leaks there are three other potential reasons for deviations from a straight line on a semi-log plot. The first is that there is residual water vapor on metal surfaces that is being de-adsorbed. Second is that there is water in the vacuum pump oil degrading its performance. The last reason is that ultimate pressure of the vacuum pump has been reached.

*3.1. Case 1: Perfect System*
A perfect system uses the assumptions of ideal gas, no pumping resistance, no external leaks into the system, and no residual entrained water vapor. The mass balance formulas are shown in equations (2) through (4)

$$\frac{dM}{dt} = \dot{m}_{In} - \dot{m}_{Out} \quad (2)$$

where $M$ = mass of gas in the system, $\dot{m}_{In}$ = mass flow rate leaking into the system, $\dot{m}_{Out}$ = mass flow rate pumping out of the system. The ideal gas law is shown in equation (3)

$$M = \frac{PV}{RT} \quad (3)$$

where $P$ = system pressure, $V$ = system volume, $R$ = gas constant, and $T$ = gas temperature. For a system with no leaks $\dot{m}_{In} = 0$. The pumping capacity is calculated per equation (4).

$$\dot{m}_{Out} = \rho Q = \frac{PQ}{RT} \quad (4)$$

where $\rho$ = system gas density and $Q$ = vacuum pump volume flow capacity. Combining (2), (3) and (4) results in equation 5

$$\frac{V}{RT}\frac{dP}{dt} = -\frac{PQ}{RT} \quad (5)$$

Equation (5) is rearranged as an integral in equation (6) and the solution to the integral is shown in equation (7). Equation (7) is rearranged to solve for $P_2$, which is the pressure at the end of pumpdown, as a function of time in equation (8)

$$\int_{P_1}^{P_2} \frac{dP}{P} = -\frac{Q}{V} \int_0^t dt \tag{6}$$

$$\ln\left(\frac{P_2}{P_1}\right) = -\frac{Q}{V} t \tag{7}$$

$$P_2 = P_1 \, e^{-\frac{Q}{V} t} = P_1 \, e^{-\frac{t}{\tau}} \tag{8}$$

where $\frac{V}{Q}$ is the time constant, $\tau$, of the system and $P_1$ is the pressure at the start of pump down

*3.2. Case 2: System with Pumping Resistance*
The second case is a system that uses the assumptions of ideal gas, pumping resistance, no external leaks into the system, and no residual entrained water vapor. The pumped mass flow rate calculation is shown in equation (9)

$$\dot{m}_{Out} = \rho_P Q = \frac{P_P Q}{RT} \tag{9}$$

where $\rho_P$ = gas density at the vacuum pump inlet, $P_P$ = pressure at the vacuum pump inlet = $P - \Delta P$, and $Q$ = vacuum pump volume flow capacity. Combining equations (2), (3) and (9) yields equation (10).

$$\frac{V}{RT} \frac{dP}{dt} = -\frac{P_P Q}{RT} = -\frac{(P - \Delta P) Q}{RT} \tag{10}$$

The pressure drop between the volume and the pump assumed to be a constant fraction of the volume pressure in order to simplify the integration into a easy-to-use analytical expression. In most cases, users will adjust the $X$ term in the calculations to match experimental data rather than trying to calculate pumping resistance using equation (11) found in engineering handbooks [2].

$$\Delta P = \frac{fL}{D} \frac{\rho V^2}{2} = \frac{fL}{D} \frac{Q^2}{2A^2} \frac{P}{RT} = XP \tag{11}$$

where $f$ = Darcy-Weisbach friction factor, $L$ = effective length of pumping line, $D$ = Inside diameter of the pumping line, $V$ = flow velocity in the pumping line = $\frac{Q}{A}$, $A$ = inside cross-sectional area of the pumping line, and the term $X = \frac{fL}{D} \frac{Q^2}{2A^2} \frac{1}{RT} = \frac{\Delta P}{P}$ used to simplify subsequent formulas. Rearranging equation (11) as an integral results in equation (12):

$$\int_{P_1}^{P_2} \frac{dP}{P(1-X)} = -\frac{Q}{V} \int_0^t dt \tag{12}$$

The solution of the integral in equation (12) assuming X term is constant is shown in equation (13). Equation (13) is rearranged to solve for $P_2$, which is the pressure at the end of pumpdown, as a function of time in equation (14).

$$\frac{\ln\left(\frac{P_2}{P_1}\right)}{1 - X} = -\frac{Q}{V} t \tag{12}$$

$$P_2 = P_1 e^{-\frac{(1-X)Q}{V}t} \tag{14}$$

### 3.3. Case 3: System with Pumping Resistance and External Leak

The third case is a system that uses the assumptions of ideal gas, pumping resistance, external leaks into the system, and no residual entrained water vapor. The flow of atmospheric air into the volume will be modeled as flow through a control valve, equating the leak as an effective control valve $C_v$. The air leak mass flow rate calculation is shown in equation (15).

$$\dot{m}_{In} = N_6 \, C_v \, Y \sqrt{F_\gamma \, X_T \, \rho_A \, P_A} \tag{15}$$

where $N_6$ = numerical constant from ISA 75.01.01 or IEC 60534, $C_v$ = equivalent valve $C_v$ representing the leak, $Y$ = gas expansion factor from ISA 75.01.01 or IEC 60534, $\rho_A$ = density of atmospheric air, $P_A$ = pressure of atmospheric air, $F_\gamma$ = specific heat ratio, $C_p/1.4$, which is 1 for air or any diatomic gas, and $X_T$ = ratio of $\Delta P / P_{In}$ for choked flow condition.

Note that $N_6' = \frac{2.73}{3600\sqrt{1000}}$ for pure SI units, $\dot{m}_{In}[=]\frac{kg}{s}$, $\rho[=]\frac{kg}{m^3}$, $P[=]Pa$. The gas expansion factor $Y$ is a function of downstream pressure until the flow becomes choked and varies from 2/3 to 1. The pump down starts with $Y=1$ and decreases until the leak is choked flow where $Y=2/3$. The pump down will spend most of the time in the choked flow regime. Since the region of interest is at low volume pressures, it is reasonable to use a constant value of $Y=2/3$. If the pressure dependence of $Y$ is considered, it significantly complicates the integration of the resulting differential equation. Assuming a value of $X_T = 0.5$ is reasonable for the inefficient leak area geometry and is consistent with ideal gas flow through an orifice without pressure recovery.

The factor G shown in equation 16 is used to simplify subsequent formulas. Note that this is a constant, which is not technically constant until the volume pressure is less than or equal to one-half of atmospheric pressure (choked flow).

$$G = N_6' \, C_v \, Y \sqrt{F_\gamma \, X_T \, \rho_A \, P_A} \tag{16}$$

Combining equations (2), (3), (9), (15) and (16) yields equation 17:

$$\frac{V}{RT}\frac{dP}{dt} = G - \frac{P\,(1-X)Q}{RT} \tag{17}$$

Rearranging equation (17) yields equation (18), which is then shown in integral form in equation (19).

$$\frac{dP}{dt} = \frac{GRT}{V} - \frac{P\,(1-X)Q}{V} \tag{18}$$

$$\int_{P_1}^{P_2} \frac{dP}{\frac{GRT}{V} - \frac{P\,(1-X)Q}{V}} = \int_{t_1}^{t_2} dt \tag{19}$$

Substituting the expressions $E = GRT/V$ and $F = (1 - X)Q/V$ into equation (19) yields equation (20)

$$\int_{P_1}^{P_2} \frac{dP}{E - PF} = \int_{t_1}^{t_2} dt \tag{20}$$

Solving the integral in equation (20) yield equation (21), which can then be rearranged to solve for the end pressure $P_2$ as a function of time as shown in equation (22)

$$\ln\left(\frac{FP_2 - E}{FP_1 - E}\right) = -Ft \tag{21}$$

$$P_2 = \frac{E + (FP_1 - E)\, e^{-\frac{(1-X)Q}{V}t}}{F} \tag{22}$$

### 3.4. Ultimate Pumping Pressure
Any leak into the system means that there is an ultimate pressure at which the pumping flow is equal to the incoming leak. It is not possible to pump below that ultimate pressure without changing the pumping system or repairing the incoming leak. This ultimate pressure can be determined by setting $\dot{m}_{In} = \dot{m}_{Out}$ and solving for $P$ as shown in equations (22) and (23). Conversely, if the ultimate pressure is known after the first pump cycle, the size of the leak can be estimated as a control valve $C_v$ equivalent.

$$\dot{m}_{In} = N'_6\, C_v\, Y\, \sqrt{F_\gamma\, X_T\, \rho_A\, P_A} = \dot{m}_{Out} = \rho(1-X)Q = \frac{P(1-X)Q}{RT} \tag{22}$$

$$P_{Ultimate} = \frac{RT}{(1-X)Q}\, N'_6\, C_v\, Y\, \sqrt{F_\gamma\, X_T\, \rho_A\, P_A} \tag{23}$$

### 3.5. Calculation Results
The plot in figure 1 uses a volume $V$ of 10 m³, a pumping capacity $Q$ of 50 m³/hr, assumed $X = \Delta P/P$ of 10%, an effective $C_v = 0.05$, and an ultimate pressure $P_{ultimate}$ of 13 mbar. As expected, Case 1 and Case 2 result in a linear line on the semi-log plot, with Case 1 having a steeper pump down slope due to the assumption of no pumping resistance. If $X$ is unknown, then $X$ can be readily fit to the experimental data at the beginning of the pumpdown. If the effective $C_v$ is unknown, then the effective $C_v$ can be readily fit to the experimental data as the pressure reaches the ultimate pressure. The region of most interest for Case 3 is below 100 mbar. The simplifying assumption was made that flow through the leak was always sonic, which is acceptable since flow through the leak is clearly in the sonic range when the volume pressure is less than 100 mbar. By the time the pressure reaches about 35 mbar the vacuum pump operator should be able to see the deviation from a straight line on a semi-log plot. After the first backfill, the operator should stop pumping at this point to keep the remaining helium in the volume as pure as possible. Prior to the first backfill it is also preferable to stop pumping before significant air ingress to keep moisture out of the volume.

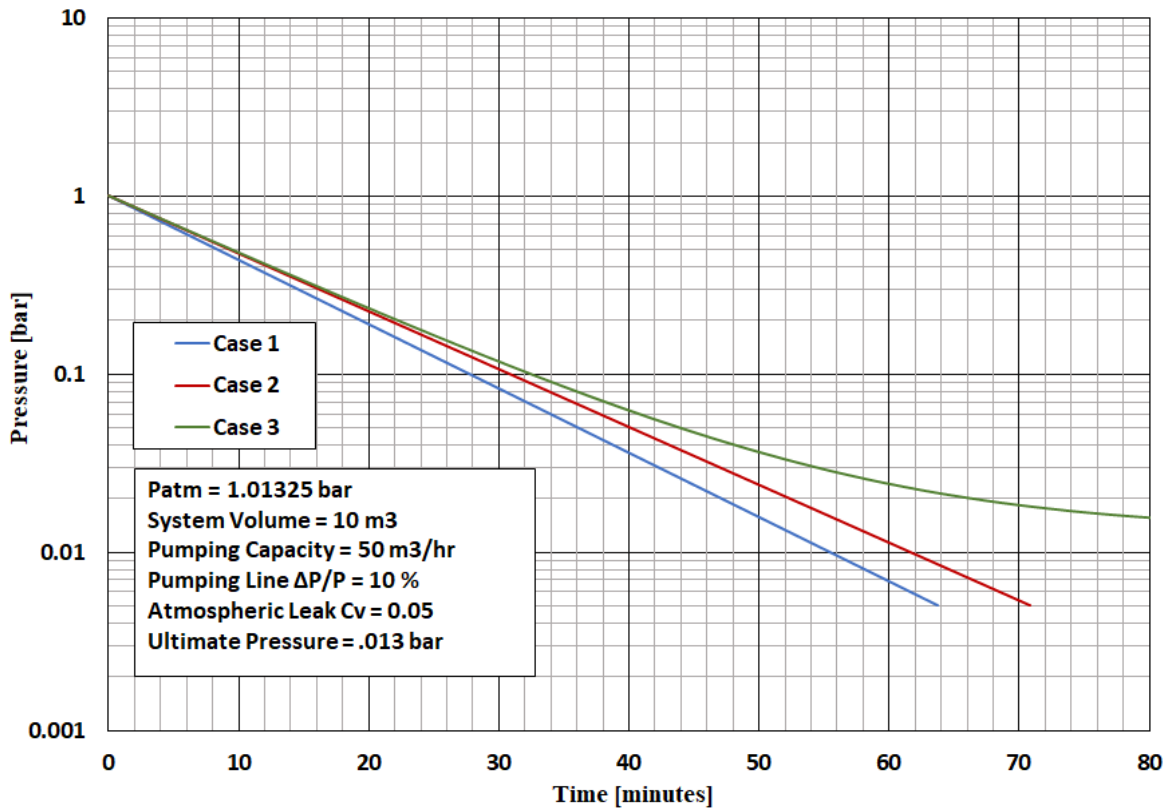

**Figure 1**. Plot showing Cases 1 through 3 as a function of time. The pumping for the Case 3 system should be cut off by 35 mbar. During the second hour of pumping the air leaking is displacing the helium-air mixture going out through the pump as the ultimate pressure is being asymptotically being approached, so the purification effect of previous pump and backfill cycles is being defeated over time.

*3.6. Experimental Results*
Large volumes and volumes which are frequently pump and backfilled should be prioritized for optimizing pump and backfill parameters in order to conserve the most helium. An obvious selection for prioritization at Fermilab was the three Vertical Test Stands used for testing bare SRF cavities. Each of these stands typically completes at least one round of pump and backfill cycles per week. Each Vertical Test Stand has 2 pressure transmitters, one in the range of 0 to 1000 mbar and the other in the range of 0 to 100 mbar to ensure accurate readings across the full pressure range of a pumpdown. The Vertical Test Stands have all metal seals since the volumes must be stringently leak tight to minimize air ingress during operation at 30 mbar while under normal testing conditions. Since very low pressures (< 3 mbar) can be readily achieved without significant deviation from a straight line on a semi-log plot, the number of pump and backfill cycles was able to be reduced to 2 cycles as shown in Figure 2. No contamination is detectable on a commercial oxygen analyzer when sweeping the Vertical Test Stand volumes after the 2 pump and backfill cycles are completed. Note that for the Vertical Test Stand a leak check is performed on the first cycle, then on the on the second cycle pumping is abruptly cutoff at around 2 mbar. Leak tightness checks should be performed before the first backfill since air leaks are only displacing air or nitrogen at that point, whereas after the first backfill any leaks are more detrimental since leaking air is displacing helium.

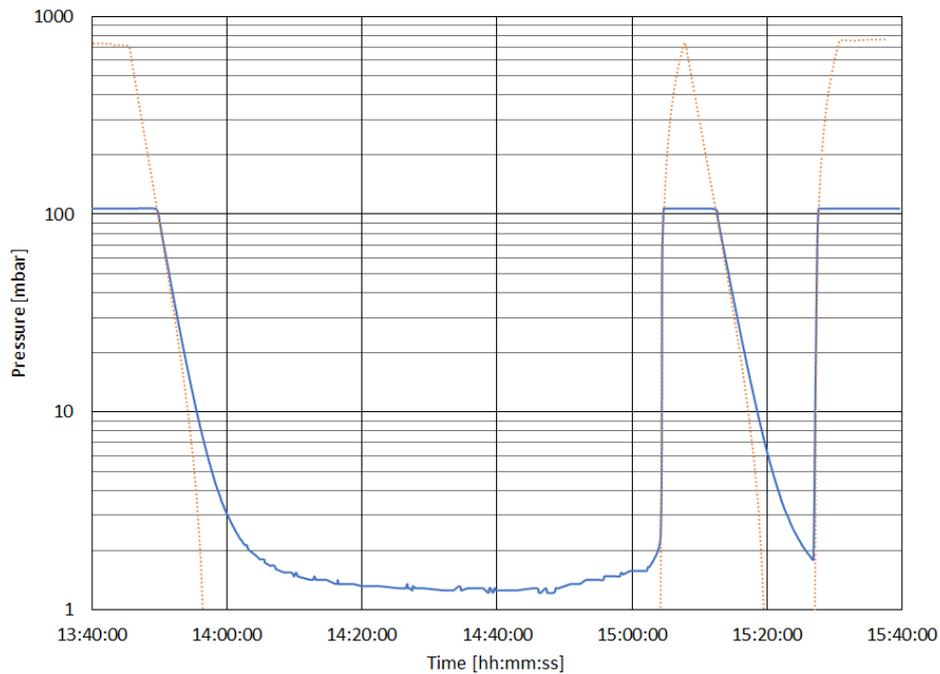

**Figure 2**. Plot of pump and backfill at VTS-3, with the solid line being the 0-100 mbar range and the dashed line being the 0-1000 mbar range.

## 4. Summary
This paper presented a methodology for optimizing pump and backfill procedures. This methodology has been verified on frequently used test stands at Fermilab. A summary of best practices for pump and backfilling is the following:
- Start by purging with dry nitrogen to remove water and get the best achievable ultimate vacuum for the system
- Use helium leak detector as part of the first pump and backfill cycle and where possible locate and repair leaks. This minimizes air in-leak during future pumping cycles and possibly reduces the number of backfills required
- Use a pressure transmitter and real-time semi-log plot to determine when to stop pumping
- Don't pump on the volume overnight during pump & backfill. Instead, have the operator watch volume pressure and cut off pumping before air ingress significantly effects the purity of the remaining helium in the volume.
- Based on the pressure when pumping is stopped, calculate the minimum number of cycles necessary to achieve desired purity level

In conclusion, it is possible to improve test schedules, reduce labor costs, and conserve helium simultaneously using optimized pump and backfill procedures.

**Acknowledgments**